\def\isarxiv{1}
\documentclass[letterpaper]{article}

\usepackage{natbib,alifeconf}  
\usepackage{url,amsmath,amssymb,textcomp}
\usepackage[colorlinks,allcolors=black]{hyperref}
\usepackage{cleveref}
\usepackage{booktabs}
\usepackage{dblfloatfix}   

\newif\ifarxiv
\ifdefined\isarxiv
  \arxivtrue
\else
  \arxivfalse
\fi

\newcommand\blfootnote[1]{%
  \begingroup
  \renewcommand\thefootnote{}\footnote{#1}%
  \addtocounter{footnote}{-1}%
  \endgroup
}

\title{Conchordal: Emergent Harmony via Direct Cognitive Coupling in a Psychoacoustic Landscape}

\author{
    Koichi Takahashi$^{1,2}$
    \mbox{}\\
    $^1$Institute for Advanced Biosciences and Graduate School of Media and Governance, Keio University \\
    $^2$Conchordal.org \\
    info@conchordal.org
} 

\begin{document}

\maketitle

\begin{abstract}
This paper introduces \textit{Conchordal}, a bio-acoustic instrument for generative composition whose sonic agents are governed by artificial life dynamics within a psychoacoustic fitness landscape. The system is built on \emph{Direct Cognitive Coupling} (DCC), a design principle requiring that generative dynamics operate directly within a landscape derived from psychoacoustic observables and read from that landscape without symbolic harmonic rules. The environment integrates roughness and harmonicity into a continuous consonance field without presupposing discrete scales or explicit harmonic rules. Agents adjust pitch through local proposal-and-accept dynamics under a crowding penalty, regulate survival via consonance-dependent metabolism, and entrain temporally through Kuramoto-style phase coupling. Four experiments are reported: (1)~consonance search produces structured polyphony with enriched consonant intervals; (2)~consonance-dependent metabolism yields survival differentials that vanish when recharge is disabled; (3)~a minimal hereditary adaptation assay shows that parent-guided respawn plus metabolic selection can accumulate more structured polyphony without adult hill-climbing; and (4)~a shared oscillatory scaffold organizes rhythmic timing under external forcing. A supplementary mechanism check reports one possible composer-configurable bridge by which spectral state can modulate temporal coupling. These findings show that a psychoacoustically derived landscape serves as an effective artificial-life terrain, yielding self-organization, selection, synchronization, and lineage-level accumulation in a non-traditional computational medium. At the level of the model, the same landscape therefore functions both as ecological terrain and as an internal proxy for musical coherence.
\end{abstract}

\ifarxiv\else
Submission type: \textbf{Full Paper}\\
\fi

Data/Code/Supplementary/Audio: \url{https://github.com/ktakahashi74/conc-paper-2026}
\ifarxiv\else
\blfootnote{\textcopyright\ 2026 Koichi Takahashi. Published under a Creative Commons Attribution 4.0 International (CC BY 4.0) license.}
\fi

\section{Introduction}

\begin{figure*}[b]
\centering
\includegraphics[width=\textwidth]{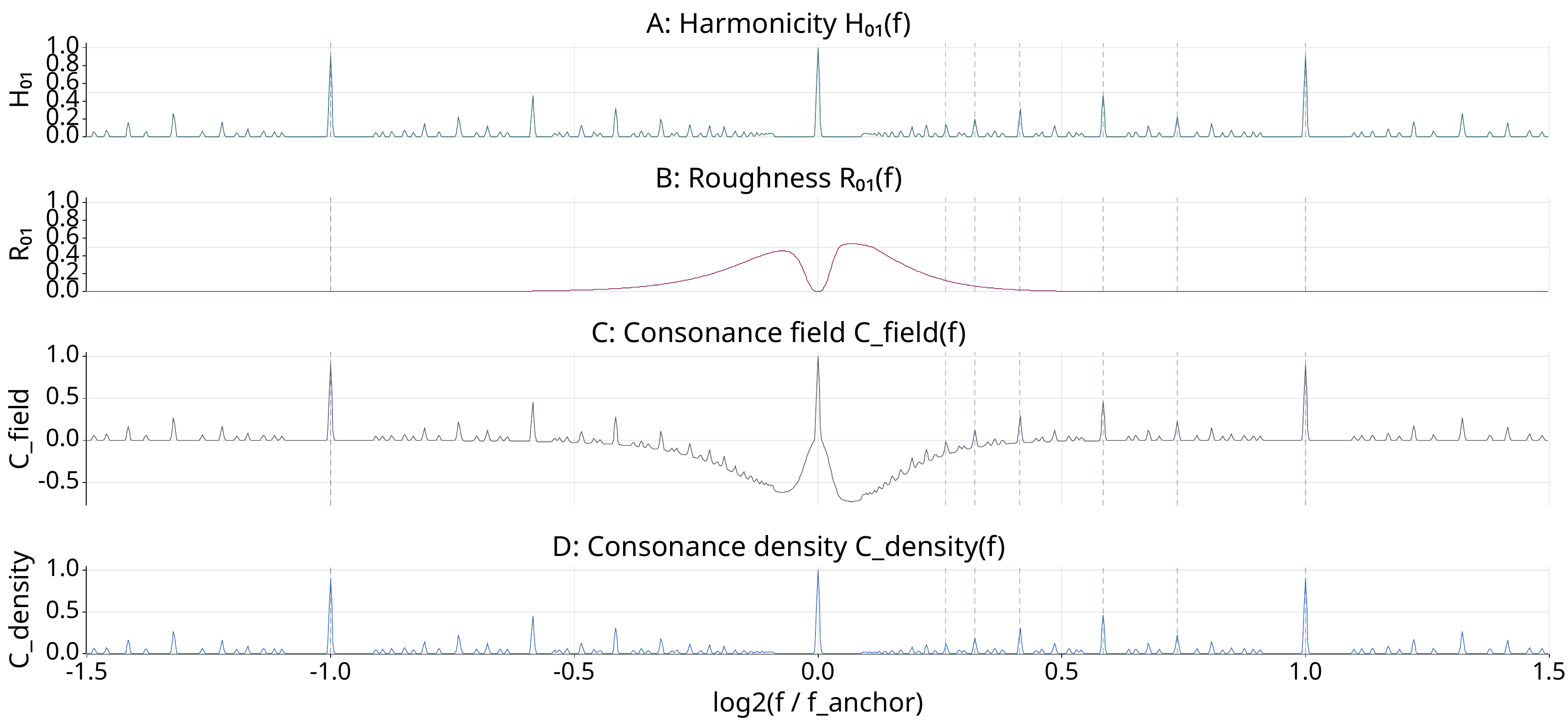}
\caption{Psychoacoustic landscape around 220\,Hz. (A)~Harmonicity $H_{01}$; (B)~roughness $R_{01}$; (C)~consonance field $C_{\text{field}}$; (D)~consonance density $C_{\text{density}}$. Grey: 0.5-oct grid; dashed: integer ratios. Simple-ratio peaks in $H$ and troughs in $R$ create attractor structure in both views. Reduced density outside $\pm 1$\,oct is a finite-order property ($N\!=\!16$).}
\label{fig:e1-landscape}
\end{figure*}

Artificial life has traditionally unfolded in spatial, chemical, or symbolic domains \citep{langton1989,bedau2000}. Yet sound---despite its rich physical structure and deep entanglement with biological perception---remains underexplored as a primary computational medium for life-like dynamics. Conchordal is a bio-acoustic instrument for generative composition that treats consonance as an ecological resource within a continuous psychoacoustic field inhabited by autonomous agents.

Conchordal is built on \emph{Direct Cognitive Coupling} (DCC), a design principle for perceptually grounded generative systems \citep{conchordal2025}. Operationally, DCC requires (i)~that the principal landscape axes be computed from perceptual observables and (ii)~that agent updates read directly from that landscape without symbolic harmonic rules. In Conchordal, cochlear roughness and periodicity-based harmonicity define a continuous consonance landscape; agents inhabit this perceptual geometry, and their ecological outcomes---survival, hereditary turnover, synchronization---are determined by landscape position. This yields an ecological--aesthetic duality: the same perceptual landscape that governs agent dynamics also supplies the system's internal proxy for musical coherence.

The scientific question of this paper is whether a psychoacoustically derived terrain can function as a non-traditional ALife substrate supporting self-organization, selection, synchronization, and hereditary turnover in controlled assays. The psychoacoustic landscape and its attractor structure are established in \S2. Four experiments test self-organization under environmental constraints, selection mediated by resource dynamics, a minimal hereditary mechanism based on lineage-biased respawn and metabolic selection, and synchronization through external forcing.

\subsection*{Related Work}

\textbf{Sonic artificial life.}
Agent-based music generation spans swarm improvisation \citep{blackwell2002}, evolutionary sonic ecosystems \citep{mccormack2003}, spectral-energy recycling \citep{eldridge2009}, rule-based autonomous composition \citep{eigenfeldt2013}, and perceptually motivated harmony navigation \citep{bernardes2016}; see \citet{miranda2002} for a survey. These systems incorporate acoustic or perceptual features into agent design, but the generative substrate---the space agents inhabit and the fitness they optimize---is not derived from a quantitative perceptual model. DCC addresses this gap.

\textbf{Computational consonance.}
Beyond the classical Plomp--Levelt roughness model \citep{plomp1965} and Terhardt's virtual-pitch framework \citep{terhardt1979}, recent work has advanced periodicity-based consonance measures \citep{stolzenburg2015} and data-driven approaches using deep neural networks \citep{harrison2020}. Conchordal adopts a first-principles approach: roughness and harmonicity are computed from cochlear and periodicity models and combined through a bilinear consonance core, keeping the landscape interpretable and free of trained parameters.

\textbf{Embodied and predictive cognition.}
DCC resonates with embodied music cognition \citep{leman2007} and predictive processing \citep{clark2013}, both grounding perception and generation in shared structures. DCC operationalises this for instrument design: agents and listeners inhabit the same perceptual geometry.

\section{The Psychoacoustic Landscape}

Unlike the tunably rugged landscapes common in artificial life \citep{wright1932,kauffman1993}, Conchordal's landscape is derived deterministically from physical acoustics and cochlear mechanics, not from arbitrary fitness functions.

\subsection{Log-Frequency Representation}

All spectral relations are computed in $x = \log_2(f)$ space, where frequency ratios become additive displacements ($\Delta x = \log_2(f_2/f_1)$). This translational invariance lets harmonic templates and interference kernels be defined as convolutions, and is essential for modelling perceptual similarity structured by integer ratios \citep{glasberg1990}. For readability, intervals and distances are reported in cents (ct); $1200$\,ct $= 1$\,octave $= 1$ unit in $x$ space ($1$\,semitone $= 100$\,ct).

\subsection{Harmonicity}

Harmonicity $H(x)$ quantifies how well spectral energy at position $x$ in $\log_2$-frequency space is explained by a harmonic template \citep{terhardt1979,parncutt1988}. Periodicity extraction---the neural basis of virtual pitch---is well localized in the auditory brainstem and primary auditory cortex, making $H$ a relatively low-level quantity (though its perceptual weighting relative to roughness varies across cultures; see Limitations).

Computation proceeds by sibling projection, a two-pass convolution with power-law weights $w_k=k^{-\rho}$ ($\rho>0$):
\begin{align}
  \mathrm{Roots}(x) &= \sum_{k=1}^{N} S(x+\log_2 k)\;k^{-\rho},
      \label{eq:h-down}\\
  H(x) &= \sum_{m=1}^{N} \mathrm{Roots}(x-\log_2 m)\;m^{-\rho}.
      \label{eq:h-up}
\end{align}
The downward pass (\ref{eq:h-down}) accumulates evidence for virtual roots; the upward pass (\ref{eq:h-up}) redistributes that evidence to harmonics. For two tones at frequency ratio $p\!:\!q$, coincident harmonics accumulate constructively; the number of such coincidences within order $N$ grows as $\max(p,q)$ shrinks, so peaks at simpler integer ratios are systematically stronger (Figure~\ref{fig:e1-landscape}A).

In all experiments $N=16$ and $\rho=0.4$. Overtone and undertone projections are blended with equal weight. The field is normalized to $H_{01}\in[0,1]$.

\subsection{Roughness}

Roughness $R(x)$ quantifies sensory dissonance from unresolved beating within the critical band \citep{plomp1965}. Like harmonicity, roughness reflects early peripheral processing: it arises from amplitude-modulation products on the basilar membrane, a mechanism rooted in cochlear mechanics \citep{glasberg1990}.

The amplitude spectrum is first mapped onto the ERB-rate scale
\begin{equation}
  z = 21.4\,\log_{10}(0.00437\,f+1)
  \label{eq:erb}
\end{equation}
and then convolved with a Plomp--Levelt interference kernel $g_{\mathrm{PL}}$ \citep{sethares1993}:
\begin{equation}
  R_{\mathrm{raw}}(z) = \int S(\zeta)\;g_{\mathrm{PL}}(|z-\zeta|)
      \,d\zeta,
  \label{eq:roughness}
\end{equation}
where $S(\zeta)$ is the spectral energy at ERB-rate $\zeta$ and $g_{\mathrm{PL}}$ peaks near $0.25$\,ERB, vanishing at both zero and wide separations. At low-order integer ratios $p\!:\!q$ (small $\max(p,q)$), the interval is wide enough that non-coincident harmonics generally fall outside each other's critical bands, so roughness troughs complement the harmonicity peaks (Figure~\ref{fig:e1-landscape}B). For higher-order ratios the interval narrows and harmonic pairs increasingly fall within the critical band, raising roughness even where harmonicity remains elevated. The raw output is normalized to $R_{01}\in[0,1]$.

\subsection{Consonance}

Unlike harmonicity and roughness, consonance is not a single well-defined perceptual quantity; it recruits higher-order circuits and is shaped by enculturation \citep{mcdermott2016}. No unique mapping $C(H,R)$ can be stipulated on purely psychoacoustic grounds. Because Conchordal uses consonance for landscape evaluation, metabolic signaling, and proposal weighting, we adopt a consonance core given by the most general bilinear form for two $[0,1]$ inputs:
\begin{equation}
C = a\,H_{01} + b\,R_{01} + c\,H_{01}\,R_{01} + d.
\label{eq:consonance-core}
\end{equation}
Here $b<0$ penalizes roughness, while the interaction term $c$ modulates the roughness penalty conditionally on harmonicity. Different coefficient settings yield task-specific quantities from the same functional family.

The evaluation field for local pitch adaptation uses $(a,b,c,d)=(1,-1.35,1,0)$:
\begin{equation}
C_{\text{field}} = H_{01} - 1.35\,R_{01} + H_{01}\,R_{01}.
\end{equation}

A second parameterisation (setting $b\!=\!0,\,c\!=\!-1$),
\begin{equation}
  C_{\text{density}} = H_{01}(1 - R_{01}),
  \label{eq:c-density}
\end{equation}
provides a nonnegative mass used in Conchordal's interactive mode for spawn-position weighting; it is not used in the experiments reported here but is shown in Figure~\ref{fig:e1-landscape}D for comparison. Figure~\ref{fig:e1-landscape}C,D illustrate the resulting landscape for a single-anchor scan; both views exhibit attractor structure near simple integer ratios (2:1, 3:2, 4:3, 5:4), consistent with classical observations on consonance \citep{helmholtz1877,sethares1993}.

For some diagnostics and supplementary bridge variants, $C_{\text{field}}$ is also passed through a sigmoid: $C_{\text{level01}} = \sigma\!\bigl(\beta(C_{\text{field}}-\theta)\bigr)\in[0,1]$, with $(\beta,\theta)=(2,0)$. This bounded map is not the direct recharge law used in the selection-pressure or hereditary assays below.

\section{Agents and Dynamics}

The Conchordal instrument is designed around an extensible agent architecture that accommodates diverse behavioural strategies---including stochastic candidate sampling, movement costs, and exploration--persistence trade-offs---without altering the underlying landscape. The experiments reported here deliberately adopt minimal dynamics to isolate the landscape's contribution to self-organisation from strategy-specific effects; each mechanism described below is a simplification of the corresponding instrument subsystem.

\subsection{Pitch Adaptation}
\label{sec:pitch-adapt}

In the reported consonance-search assay, each agent proposes local pitch perturbations of magnitude $\lambda=25$\,ct sampled within a bounded neighborhood in log-frequency space. A hill-climbing acceptance rule evaluates candidate positions via $C_{\text{field}}$ and a crowding penalty computed in ERB space. The crowding kernel $K_{\text{crowd}}(\Delta_{\text{erb}})$ is derived analytically as the complement of the roughness kernel: $K_{\text{crowd}}(d) = \max\!\bigl(0,\; 1 - R(d/\kappa) / R_{\text{peak}}\bigr)$, where $R$ is the full Plomp--Levelt roughness kernel and $\kappa$ is a reach factor ($\kappa=2.1$). This construction repels agents that are too close in pitch---preventing collapse to unison---while yielding to landscape-mediated repulsion at wider separations where roughness peaks. The combined score is $S = C_{\text{field}} - \lambda_C \sum_{j \neq i} K_{\text{crowd}}(\Delta_{ij}) - \lambda_M |\Delta x|$, where $\lambda_M$ is a move-cost coefficient in log-frequency space. Proposals are accepted with probability $\min\!\bigl(1,\,\exp\bigl(\Delta S/T\bigr)\bigr)$ where $\Delta S = S_{\text{new}} - S_{\text{old}}$ and $T$ decays exponentially from $T_0=0.05$ with $\tau=4.8$ sweeps, resetting at the phase switch.

\subsection{Metabolic Policy}

Agents possess energy $E$. In the selection-pressure assay, dynamics follow:

\begin{equation}
E \leftarrow E - c_b \Delta t + r_E \cdot \max(0, C_{\text{score}}) \cdot \Delta t,
\end{equation}

where $c_b$ is basal cost and $r_E$ is a consonance-dependent recharge rate. Only positive raw $C_{\text{score}}$ from the reference landscape contributes directly to recharge; the bounded statistic $C_{\text{level01}}$ is recorded for analysis but is not itself the recharge law. Agents start with $E=1.0$ and perish when $E\leq 0$. We set $c_b=0.5$/s and $r_E=0.4$/s; time advances in hops of $512$ samples at $48\,$kHz ($\Delta t\approx 10.7\,$ms). In the baseline condition, recharge scales continuously with positive consonance ($r_E=0.4$/s); in the ablation condition, $r_E=0$. In the hereditary assay, the lifecycle is reset to a lower initial energy and slower basal drain ($E_0=0.05$, $c_b=0.12$/s), and the contextual score is first mapped to a bounded survival signal,
\[
s_i = \mathrm{clamp}\!\bigl((C_{\text{score},i}-0.30)/(0.80-0.30),\,0,\,1\bigr),
\]
and recharge then uses $s_i$ rather than raw $C_{\text{score}}$, with $r_E=0.20$/s in the selection-on hereditary runs. An agent's \emph{vitality} $v = \sqrt{E/E_{\text{cap}}}$ (where $E_{\text{cap}}=1.0$) provides a normalised vigour signal available to downstream dynamics such as phase coupling.

\subsection{Temporal Oscillation}

Neural oscillations in auditory cortex entrain to periodic acoustic stimuli across multiple frequency bands \citep{lakatos2008,doelling2015}, and computational models of beat tracking posit coupled-oscillator dynamics as the underlying mechanism \citep{large1999}. Conchordal's full implementation realises this temporal axis of DCC through a multi-band modulation bank spanning delta through beta frequencies ($\approx$0.5--30\,Hz), mirroring layered temporal processing in cortical auditory networks. Agents couple to this shared oscillatory field rather than directly to one another (mean-field approximation), and the interaction is bidirectional: the field entrains agent timing while agent onsets excite the field.

The simplified experiments isolate the core coupling law as a single phase oscillator per agent driven toward a shared phase reference:
\begin{equation}
\dot{\phi}_i = \omega_i + K\sin\!\bigl(\theta_i^{\ast}(t) - \phi_i\bigr),
\end{equation}
where $K$ is coupling strength and $\theta_i^{\ast}(t)$ is the condition-specific drive phase: shared in the baseline scaffold, cycle-reset in the scrambled condition, and effectively disabled when the scaffold is off. Neural entrainment motivates the temporal axis itself, but current neuroscience does not uniquely determine how spectral state should modulate temporal coupling: harmonicity, attention, predictive context, and higher-order control all provide plausible links. Conchordal therefore treats spectral-to-temporal bridge mappings as composer-configurable design choices rather than fixed neurocognitive laws. The temporal scaffold assay below tests the shared oscillatory field directly; one vitality-gated bridge used in the current implementation is documented separately as a supplementary mechanism check. Phase locking is quantified via the Kuramoto order parameter $R(t)$ \citep{kuramoto1984,strogatz2000} and phase-locking value (PLV) \citep{lachaux1999}.

\section{Results}

\begin{figure*}[t]
\centering
\includegraphics[width=\textwidth,trim=56 0 56 0,clip]{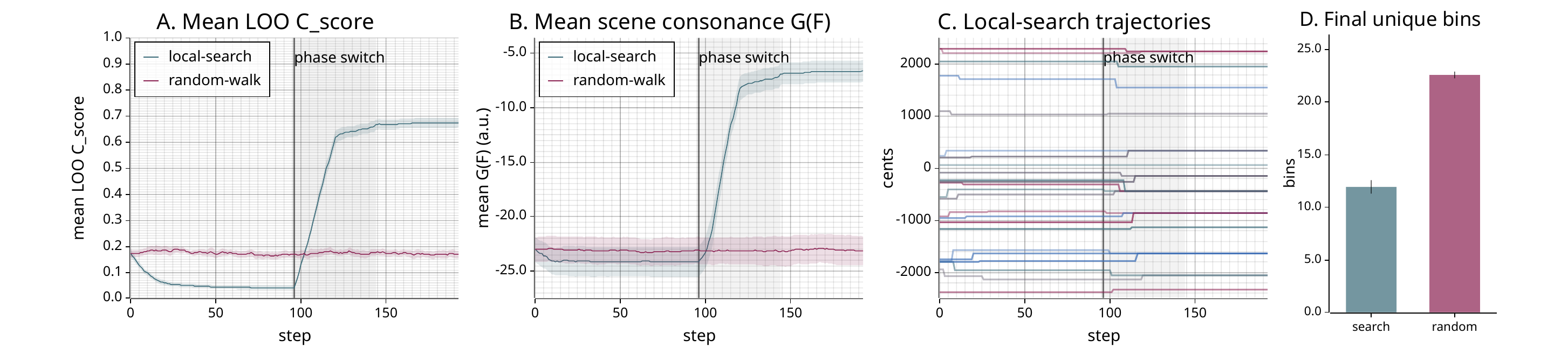}
\vspace{0.3em}
\includegraphics[width=\textwidth]{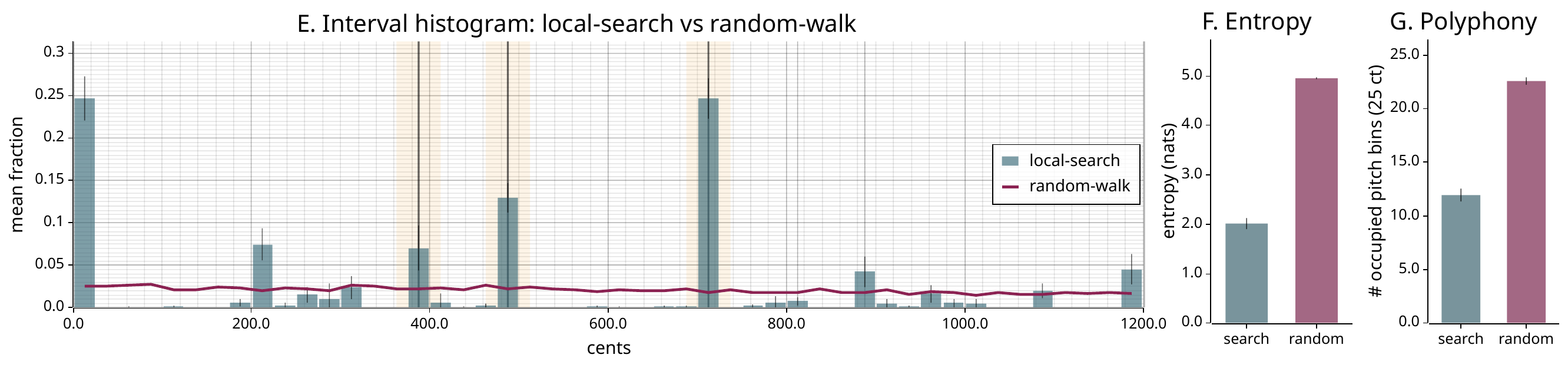}
\caption{Consonance search. \emph{Top}---(A)~Mean LOO consonance $C_{\text{score}}$ over time (95\% CI); vertical line = phase switch, shaded band = burn-in excluded from statistics. (B)~Scene consonance $G(F)$ (Eq.~\ref{eq:consonance-core}) after singleton subtraction (95\% CI). (C)~Representative local-search trajectories (cents from drone). (D)~Final unique pitch bins (95\% CI). \emph{Bottom}---(E)~Pairwise interval histogram (25\,ct bins): \emph{local-search} concentrates mass at consonant intervals; \emph{random-walk} spreads broadly. (F)~Entropy of the 5\,ct interval distribution (95\% CI).}
\label{fig:e2-dynamics}
\label{fig:e2-intervals}
\end{figure*}

\begin{figure*}[t]
\centering
\includegraphics[width=\textwidth]{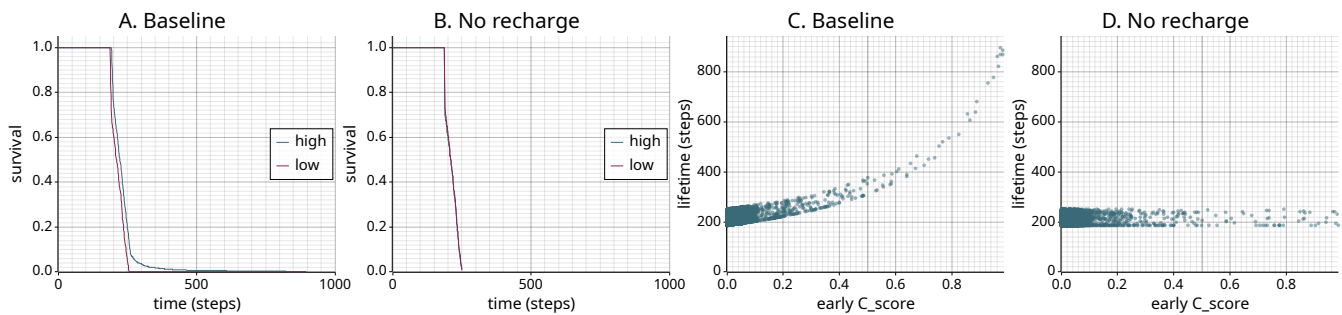}
\caption{Consonance as selection pressure. (A,\,B) Kaplan--Meier survival by early consonance ($C_{\mathrm{firstK}}$) median split: with recharge, high-consonance agents survive longer; without recharge, the separation disappears. (C,\,D) Lifetime versus early consonance shows the same contrast. Curves pool events across 20 seeds for display; inference is seed-level ($n=20$).}
\label{fig:e3-selection}
\end{figure*}

Four experiments each isolate a different mechanism subset (Table~\ref{tab:conditions}). All $\pm$ intervals denote $\pm 1$\,SD across seeds unless otherwise noted.

\begin{table}[t]
\centering
\caption{Experiment conditions. \checkmark/\texttimes\ = on/off; abl.\ = ablated (compared with both states). Results are discussed in the order consonance search, selection, hereditary adaptation, and temporal scaffolding. All experiments use a fixed 220\,Hz reference; in the consonance-search assay this is realised as a fixed drone.}
\label{tab:conditions}
\footnotesize
{\setlength{\tabcolsep}{4pt}
\begin{tabular}{@{}lcccc@{}}
\toprule
 & \textbf{Search} & \textbf{Selection} & \textbf{Heredity} & \textbf{Entrainment} \\
\midrule
Main figure & Fig.~\ref{fig:e2-dynamics} & Fig.~\ref{fig:e3-selection} & Fig.~\ref{fig:exp4-hereditary} & Fig.~\ref{fig:e7-temporal} \\
$N$ agents  & 24  & 32  & 16  & 32 \\
Seeds       & 20  & 20  & 20  & 20 \\
Hill climb  & abl.\ & \texttimes & \texttimes & \texttimes \\
Crowding    & \checkmark & \texttimes & \checkmark & \texttimes \\
Metabolism  & \texttimes & abl.\ & abl.\ & \texttimes \\
Entrainment & \texttimes & \texttimes & \texttimes & \checkmark \\
Heredity    & \texttimes & \texttimes & abl.\ & \texttimes \\
Duration    & 8 sweeps & 200 deaths & \shortstack{12k steps\\2.5k deaths} & \raisebox{0.4ex}{40\,s} \\
\bottomrule
\end{tabular}
}
\end{table}

\subsection{Consonance Search}

To test anchored harmony generation around a fixed reference, twenty-four adaptive voices plus a fixed drone at 220\,Hz are placed on a discrete log-frequency grid (3\,ct bins) spanning $\pm 2$ octaves from the drone. In the \emph{local-search} condition, agents generate local pitch proposals and accept them according to an exact leave-one-out (LOO) objective combining consonance with a crowding penalty. The \emph{random-walk} ablation keeps the same schedule, step window, and pitch bounds but replaces hill-climbing with a matched random local walk. LOO consonance is evaluated by removing each agent's spectral contribution from the shared environment before rescoring---an exact LOO that the real-time instrument approximates with a lighter-weight harmonic-projection subtraction. The LOO mean of $C_{\text{field}}$ across all agents defines $C_{\text{score}}$, the primary outcome metric and the quantity that drives local updates. By contrast, \(G(F)\) is used only as a post-hoc scene-level diagnostic of population-wide interaction structure.

Each run consists of 8 sweeps: a preparatory dispersion phase (steps 0--3) followed by a consonance-seeking phase (steps 4--7), with the first 2 sweeps treated as burn-in in the statistical summaries. The pre-switch phase negates the same exact-LOO proposal score ($S \to -S$), so under the shared crowding term agents leave locally consonant crowded basins and broaden basin coverage before final convergence. A consonance-only control (Supplementary S7) shows that structure also emerges without this curriculum, but with fewer stable pitch positions ($8.4$ vs.\ $11.9$ unique bins; $p < 0.001$) and tighter pitch packing, so the curriculum's benefit is primarily spatial rather than a simple gain in endpoint $C_{\text{score}}$.

\textbf{Diversity.} Unique pitch bins averaged $11.95 \pm 1.40$ (\emph{local-search}) vs.\ $22.60 \pm 0.80$ (\emph{random-walk}). Under the crowding penalty and two-phase curriculum, hill-climbing concentrates 24 agents into $\approx$\,12 distinct pitch classes at consonant intervals, whereas random walks distribute them nearly uniformly across the grid. Scene consonance $G(F)=aH_{\mathrm{soc}}+bR_{\mathrm{soc}}+cH_{\mathrm{soc}}R_{\mathrm{soc}}$ (Eq.~\ref{eq:consonance-core}), where $H_{\mathrm{soc}}$ and $R_{\mathrm{soc}}$ denote scene harmonicity and roughness after subtracting the mean singleton contribution, shows that the major global gain occurs after the phase switch for \emph{local-search}, while \emph{random-walk} remains flat (Figure~\ref{fig:e2-dynamics}B).

\textbf{Interval Structure.} Shannon entropy of the pairwise interval distribution (240 bins of 5\,ct over 0--1200\,ct; uniform maximum $\ln 240 = 5.48$\,nats) was $2.013 \pm 0.257$\,nats (\emph{local-search}) vs.\ $4.961 \pm 0.049$ (\emph{random-walk}). The \emph{local-search} distribution concentrates mass at a small number of just-intonation intervals (Figure~\ref{fig:e2-dynamics}E), whereas \emph{random-walk} approaches a near-uniform distribution. An independent just-intonation proximity score ($\mathrm{JI}_{\text{score}}$, computed as weighted proximity to simple ratios with $p,q \leq 8$) confirms the result: \emph{local-search} $0.366 \pm 0.051$ vs.\ \emph{random-walk} $0.115 \pm 0.013$.

These results show that hill-climbing is the primary source of structured polyphony: replacing it with matched random-walk updates disperses agents broadly and raises interval entropy toward the uniform ceiling, a pattern reinforced by the shuffled-landscape control (\S Terrain Validity).

\subsection{Consonance as Selection Pressure}

Each run maintained agents within $\pm 1$ octave of the 220\,Hz anchor; dead agents were immediately respawned at random positions until 200 deaths accumulated ($\approx$4{,}000 pooled lifetimes per condition across 20 seeds). Early consonance $C_{\mathrm{firstK}}$ (mean $C_{\text{level01}}$ over the first $K\!=\!20$ ticks) strongly predicted lifetime under baseline recharge (Figure~\ref{fig:e3-selection}). The primary analysis operates at the seed level to avoid inflated significance from pooling non-independent lifetimes within a run: seed-level analysis ($n\!=\!20$, Fisher $z$) yielded mean Pearson $r = 0.718$, corresponding to a raw mean $r = 0.704 \pm 0.100$ (95\% CI $[0.67, 0.76]$; $p < 0.001$). Pooled ($n\!\approx\!4{,}000$) results are consistent: $r = 0.763$; median survival 222 vs.\ 210.

When recharge was disabled, seed-level mean $r = 0.000 \pm 0.068$, not significantly different from zero; the condition contrast was significant (Welch $t(23.2) = 18.7$, $p < 0.001$, Fisher $z$). Because pitch adaptation is disabled (Table~\ref{tab:conditions}), early consonance is determined entirely by initial position; the ablation confirms that the survival differential requires consonance-dependent recharge, not positional coincidence. Consonance thus functions as an ecological resource, generating selection pressure through the metabolic pathway.

\begin{figure*}[!t]
\centering
\includegraphics[width=\textwidth]{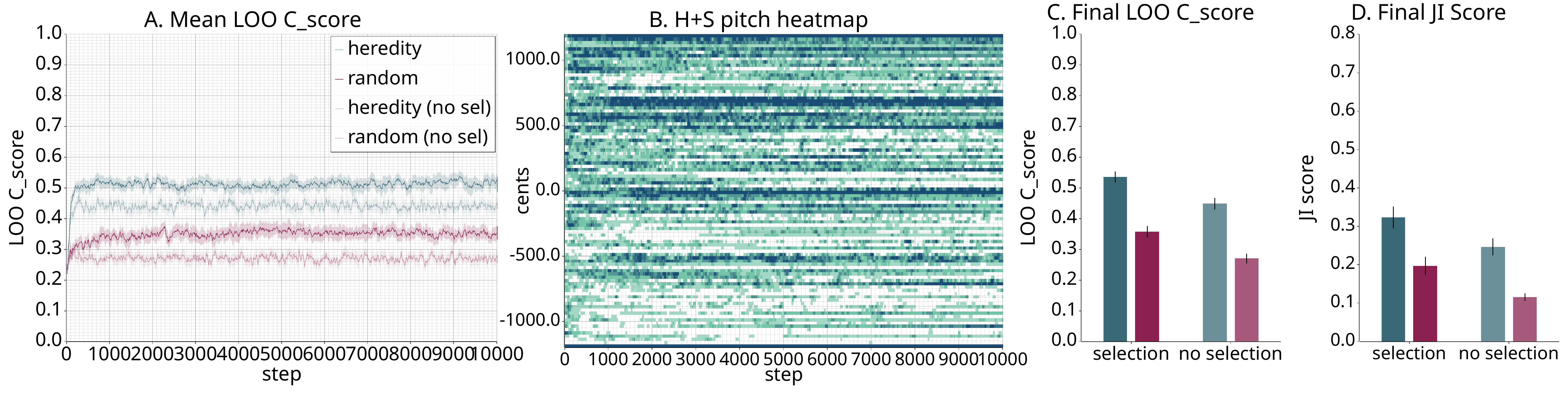}
\caption{Hereditary adaptation through lineage-biased respawn and metabolic selection (20 matched seeds, 4 conditions). (A)~Mean leave-one-out contextual $C_{\text{score}}$ over simulation steps (95\% CI). (B)~Pitch heatmap for heredity$+$selection over simulation steps. (C)~Final leave-one-out contextual $C_{\text{score}}$ by condition (95\% CI). (D)~Final just-intonation proximity score by condition (95\% CI). Teal denotes heredity and rose denotes matched random; dark colors indicate selection-on and light colors indicate no-selection.}
\label{fig:exp4-hereditary}
\end{figure*}

\subsection{Minimal Hereditary Adaptation}
\label{sec:exp4-heredity}

The first two assays establish structure and selection but not heredity, so the minimal hereditary assay tests lineage-level accumulation directly. The factorial remains $2\times 2$: \emph{respawn} (heredity vs.\ matched log-random baseline) $\times$ \emph{metabolic selection} (on vs.\ off), with adult pitch locked in all conditions. Newborns receive only a short slot-local settling phase (2 ticks), so the main difference is whether respawn is guided by parent-biased peak choice or by log-random candidate sampling under the same scene filter. In the heredity arm, family/slot choice is lineage-biased but the final azimuth within the chosen family is selected by local search rather than inherited directly, because direct azimuth inheritance systematically missed the sharp local maxima of the consonance landscape.

In this assay, heredity$+$selection is clearly best on all four musical summaries in the main comparison (Figure~\ref{fig:exp4-hereditary}). Final leave-one-out contextual $C_{\text{score}}$ rises to $0.536 \pm 0.039$ for heredity$+$selection, versus $0.358 \pm 0.044$ for matched random$+$selection; interval entropy falls from $4.220 \pm 0.060$ to $3.281 \pm 0.098$; the just-intonation proximity score rises from $0.197$ to $0.323$; and the number of occupied pitch bins contracts from $14.35$ to $12.30$, indicating structured concentration rather than diffuse scattering. The heredity-only arm also outperforms matched random without selection ($C_{\text{score}}$ $0.449 \pm 0.043$ vs.\ $0.271 \pm 0.036$), showing that lineage bias itself already contains useful information about where musically viable offspring can be placed.

The time series in Figure~\ref{fig:exp4-hereditary}A show that the effect is not only an endpoint difference: the heredity$+$selection arm reaches a much higher mean trajectory area ($\mathrm{AUC}_C = 0.510$ vs.\ $0.350$ for matched random$+$selection), and 15 of 20 seeds cross the $\mathrm{JI}_{\text{score}} \ge 0.50$ threshold whereas none of the matched-random runs do. Informal listening to the replay renders agrees: only heredity$+$selection produces clearly musical multi-band polyphony, while the controls remain more diffuse. An unrestricted hard-random restart baseline can still reach higher raw $C_{\text{score}}$ in supplementary runs, but it does so by collapsing into very few occupied bands ($5.75$ bins on average), so it is not an adequate control for musical polyphony. Thus heredity matters when it guides which harmonic family is revisited, but the final fine pitch within that family must still be re-optimized locally.

\begin{figure*}[t]
\centering
\includegraphics[width=\textwidth]{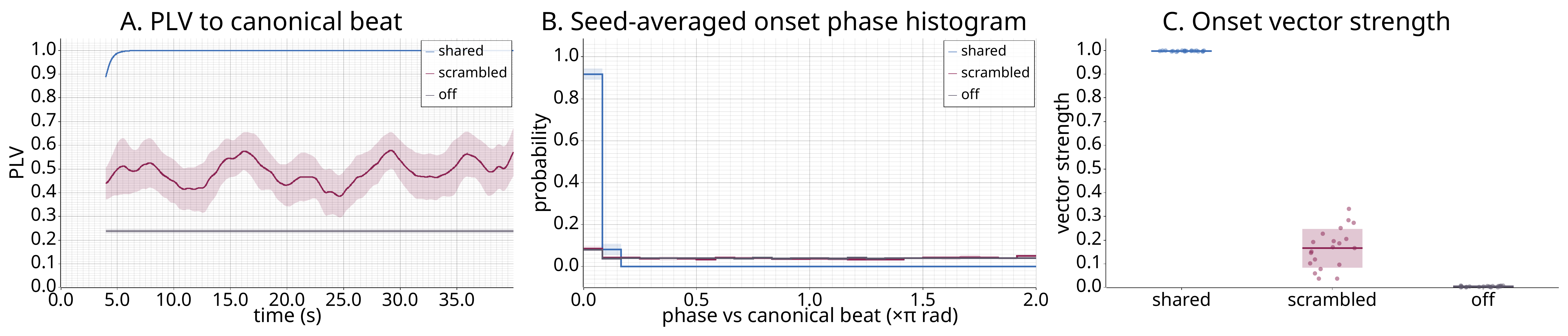}
\caption{Temporal scaffold assay (20 seeds $\times$ 3 conditions). (A)~Group mean PLV to the canonical 2\,Hz (120 BPM) beat over time (95\% CI). (B)~Seed-averaged onset phase histogram relative to the canonical beat (bin probabilities normalized within seed, then averaged across seeds; shaded bands indicate 95\% CI across seeds). (C)~Seed-level onset vector strength: shared rhythm produces the strongest phase concentration, scrambled forcing is weaker, and off is lowest.}
\label{fig:e7-temporal}
\end{figure*}

\subsection{Shared Rhythm Scaffold Organizes Timing}

The temporal scaffold assay isolates the temporal axis of DCC. Rather than coupling synchrony to consonance or metabolic state, all agents are reduced to phase oscillators driven by an external 2\,Hz (120 BPM) scaffold with fixed coupling strength ($K_{\mathrm{base}}\!=\!3.0$; intrinsic frequency $1.8\!\cdot\!2\pi$\,rad/s, $\pm 2\%$ jitter; 40\,s runs, 20 seeds each). This removes spectral confounds and tests whether a shared oscillatory field organizes timing, and whether disrupting that shared temporal reference weakens phase concentration.

Three conditions were compared. In \emph{shared}, all agents couple to the same continuous 2\,Hz phase reference. In \emph{scrambled}, the drive retains the same 2\,Hz cycle length and coupling magnitude, but its phase is re-randomized at each cycle boundary, destroying cross-cycle temporal continuity while preserving periodic energy. In \emph{off}, external drive coupling is disabled ($k_{\text{eff}} = 0$); agents retain intrinsic frequencies but receive no periodic forcing. All statistics are evaluated against the same fixed 2\,Hz reference beat, so the comparison tests stable alignment to a shared beat rather than momentary tracking of each condition's own drive.

Figure~\ref{fig:e7-temporal} shows the expected ordering throughout: shared forcing yields the strongest PLV and phase concentration, scrambled is weaker, and off is lowest. Seed-level onset vector strength is $0.998 \pm 0.002$ for shared, $0.166 \pm 0.083$ for scrambled, and $0.005 \pm 0.003$ for off (all pairwise Welch $t \geq 8.7$, $p < 0.001$). This assay validates that Conchordal's temporal axis can support controlled beat alignment. The result is expected under coupled-oscillator theory and is included as temporal-axis validation rather than as a claim of spontaneous rhythmic emergence from consonance alone.

\section{Discussion}

\subsection{Integrated Interpretation}

The four experiments form a cumulative argument: consonance search shows structured configurations beyond trivial collapse, metabolic selection turns consonance into a survival differential, minimal hereditary adaptation accumulates this structure across generations when lineage bias guides respawn, and temporal scaffolding organizes timing. Together they show that a psychoacoustic landscape can sustain the core ingredients of an artificial ecology without symbolic encoding. A supplementary mechanism check (Supplementary S3.1) then documents one possible spectral--temporal bridge in the current implementation without claiming it as a uniquely validated law. In the hereditary assay (Figure~\ref{fig:exp4-hereditary}), the successful combination is specifically family heredity plus local azimuth search plus metabolic selection: heredity alone is informative, selection alone is insufficient under the matched baseline, and direct azimuth inheritance is worse than re-optimizing fine pitch locally. This ties the hereditary claim back to the sharply peaked structure of the consonance landscape.

\subsection{The Ecological--Aesthetic Duality}

In conventional generative music systems, ecological fitness and aesthetic evaluation are separate functions. DCC proposes an ecological--aesthetic duality: because Conchordal's terrain is computed directly from psychoacoustic observables, the same landscape that governs survival, movement, and hereditary turnover also supplies the system's internal proxy for musical coherence. This is a claim about model structure, not yet a claim about validated human judgement. The four assays support that structural claim in complementary ways: consonance search shows that landscape-following dynamics yield organized polyphony rather than arbitrary dispersion; metabolic selection makes higher-consonance states ecologically advantageous; hereditary adaptation accumulates those states across generations; and temporal scaffolding validates the temporal axis under controlled forcing. Terrain controls (\S Terrain Validity) further show that this coupling depends on coherent psychoacoustic structure rather than merely a scalar reward, while roughness and crowding penalties prevent collapse to the ecologically easy but musically trivial solution of unison.

\subsection{Terrain Validity}

Because all in-simulation metrics derive from the same psychoacoustic model that defines the fitness landscape, a central question is whether the reported structure depends on the specific properties of that landscape or would arise under any scalar fitness field. To address this circularity, we introduced matched terrain controls (Supplementary~S8).
In a shuffled-landscape control, a fixed random permutation of the consonance-score bins---generated once per run and held fixed throughout all sweeps---was applied, preserving the marginal score distribution while destroying spatial coherence.
Pitch spread did not collapse under shuffling; instead, crowding alone over-dispersed the population ($11.9$ vs.\ $17.0$ unique bins, $p < 0.001$; nearest-neighbour spacing $0.576$ vs.\ $0.928$\,ct, $p < 0.001$), while interval structure degraded substantially: entropy rose from $2.013 \pm 0.257$ to $4.416 \pm 0.168$\,nats ($p < 0.001$).
Thus, it is not merely the presence of a fitness signal that matters; a coherent psychoacoustic gradient is required for the emergence of structured polyphony.
The current hereditary adaptation assay has not yet been rerun through the full shuffled-landscape and topology-variation battery used for the consonance-search assay, so the hereditary claim should presently be read as mechanism-level evidence within one well-characterized terrain rather than as a fully general terrain-invariance result. Extending the same terrain-validity battery to the hereditary assay is the next robustness check.

\section{Limitations and Future Work}
\label{sec:limits}

The ecological--aesthetic duality predicts that ecological success correlates with perceptual coherence, but the present work does not yet provide independent behavioural validation. Because harmonicity--roughness weighting varies across listeners and cultures \citep{mcdermott2016}, the present coefficients implicitly encode a Western-leaning profile; listener-informed reparametrisation could enable non-Western perceptual weightings. All experiments use a fixed 220\,Hz reference; in the consonance-search assay this reference is realised as a fixed drone. Including a fixed point is a compositional choice, not a system constraint: conventional instruments fix intonation at construction, whereas adaptive agents traverse pitch space continuously, so that removing the reference allows the emergent tuning itself to evolve during performance. Experiments with freely drifting populations and inharmonic timbres would test whether the present findings generalise to such conditions.

Ensemble size introduces a further capacity constraint: pairwise spectral conflicts grow approximately as $O(N^2)$, whereas the number of distinct high-consonance basins grows much more slowly. Anti-fusion mechanisms are therefore needed to avoid trivial unison/octave collapse, but dense polyphony and controlled dissonance can still be musically valuable even when they lower the current score. This leaves room to refine the consonance quantity, making $N$ and crowding strength joint ecological/compositional control parameters rather than neutral scaling choices.

The hereditary adaptation assay converges to fixed-landscape attractors and does not constitute open-ended evolution \citep{bedau2000}; whether evolvable timbres or coupling topologies could support open-ended dynamics remains open. More broadly, closing the loop via real-time listener biosignals---feeding back brainwave or physiological signals to the landscape---would realise the bidirectional DCC envisioned in the Conchordal manifesto, turning perception and generation into a coupled dynamical system.

\section{Conclusion}

Conchordal demonstrates that a continuous landscape derived from psychoacoustic observables---harmonicity and roughness---can sustain the core ingredients of artificial ecology: self-organisation, selection, entrainment, and a minimal hereditary mechanism. Terrain controls directly confirm this dependence for the consonance-search assay, while the hereditary result currently remains mechanism-level evidence within one well-characterized terrain. Taken together, these results establish DCC-derived consonance as a viable ALife substrate that complements the spatial and chemical media traditionally studied in the field.

A central consequence is an ecological--aesthetic duality: because the fitness landscape is derived from psychoacoustic models of auditory processing, the same landscape governs ecological dynamics and supplies the system's internal proxy for musical coherence. As an instrument, Conchordal thus offers composers a generative substrate in which ecological dynamics produce musically coherent material without symbolic specification. More broadly, DCC is not inherently auditory: any creative medium whose perceptual basis admits quantitative modelling---visual, tactile, or kinesthetic---could host an analogous ecology.

\section{Acknowledgments}

The author thanks Gilberto Bernardes for helpful correspondence regarding the naming of \textit{Conchordal} in relation to his \textit{Conchord} system \citep{bernardes2016}.


\footnotesize
\bibliographystyle{apalike}
\bibliography{conc}

\end{document}